\documentclass[aps,a4,epsf,8pt,twocolumn,showpacs]{revtex4}

\usepackage{amsmath}
\usepackage{amssymb}
\usepackage{graphicx}
\usepackage[usenames]{color}

\begin{document}

\newcommand{\blue}[1]{\textcolor{blue}{#1}}
\newcommand{\beq}{\begin{equation}}
\newcommand{\eeq}{\end{equation}}
\newcommand{\beqa}{\begin{eqnarray}}
\newcommand{\eeqa}{\end{eqnarray}}
\newcommand{\bmat}{\begin{displaymath}}
\newcommand{\emat}{\end{displaymath}}

\newcommand{\eq}[1]{Eq.~(\ref{#1})}

\newcommand{\lan}{\langle}
\newcommand{\ran}{\rangle}

\title{Spin-chirality decoupling and critical properties of a two-dimensional fully frustrated {\it XY\/} model}

\author{Soichirou Okumura, Hajime Yoshino and Hikaru Kawamura}
\affiliation{Department of Earth and Space Science, Faculty of Science,
 Osaka University, Toyonaka 560-0043, Japan
}

\begin{abstract}
We study the ordering of the spin and the chirality in the fully frustrated {\it XY\/} model on a square lattice by extensive Monte Carlo simulations. Our results  indicate unambiguously that the spin and the chirality exhibit separate phase  transitions at two distinct temperatures, {\it i.e.\/}, the occurrence of the spin-chirality decoupling. The chirality exhibits a long-range order at $T_{\rm c}=0.45324(1)$ via a second-order phase transition, where the spin remains disordered with a finite correlation  length $\xi_{\rm s}(T_{\rm c}) \sim 120$. The critical properties of the chiral transition determined from a finite-size scaling analysis for large enough systems of linear size $L > \xi_{\rm s}(T_{\rm c})$ are well compatible with the Ising universality. On the other hand, the spin exhibits  a phase transition at a lower temperature $T_{\rm s}=0.4418(5)$ into the quasi-long-range-ordered phase. We found $\eta(T_{\rm s})=0.201(1)$, suggesting that the universality of the spin transition is different from that of the conventional Kosterlitz-Thouless (KT) transition.
\end{abstract}

\pacs{75.10.Hk,74.81.Fa,74.25.Qt}

\date{\today}
\maketitle

\section{Introduction}

 Fully frustrated {\it XY\/} (FFXY) model is a basic statistical mechanical model of geometrically frustrated systems \cite{Villain,TJ83}. An example is the antiferromagnetic {\it XY\/} spin system on the two-dimensional (2D) triangular lattice \cite{MS84}. Another realization is a Josephson junction array (JJA) on the square lattice under an external magnetic field such that the flux density per plaquette is $f=1/2$ \cite{Mooij-group,Martinoli-Leemann}. These systems possess an Ising-like $Z_{2}$ symmetry with respect to global spin reflections in spin space and a continuous $U(1)$ symmetry with respect to global spin rotations in spin space. 

 The ground state of the model is characterized by the opposite sense of the chirality. The chirality is an Ising-like variable taking either plus or minus corresponding to the right- and left-handed noncollinear spin structure \cite{Villain}. More precisely, in the ground state of the model, the chirality exhibits a checker-board-like pattern much as in the antiferromagnetic Ising model on the square lattice. Since the chirality changes its sign with respect to global spin reflections in spin space, the $Z_{2}$ spin-reflection symmetry is broken in the ground state of the model. Then, by an analogy to the 2D Ising model, it would be natural to expect that the ordering of the chirality takes place at a finite critical temperature $T_{\rm c}$.  Concerning the continuous $U(1)$ spin symmetry, although the Mermin-Wagner theorem inhibits the onset of a true spin long-range order at any finite temperature \cite{Mermin-Wagner}, a finite-temperature  transition into a quasi-long-range-ordered state like the Kosterlitz-Thouless (KT) transition \cite{KT} is still expected to occur.

 A central issue in the studies of the FFXY model, since the seminal work by  Teitel and Jayaprakash \cite{TJ83},  has been to clarify how these two distinct types of orderings take place. An interesting possibility among others, which has attracted researchers for decades, is that even at the temperature where the chirality establishes a long-range order the spin (superconducting phase in case of the JJA) may remain disordered due to thermally excited, unbound vortices. Then, the orderings of the chirality and of the spin take place at two separate temperatures $T=T_{\rm c}$ and $T=T_{\rm s}$ such that $T_{\rm c} > T_{\rm s}$. We emphasize that the possibility of such a ``spin-chirality decoupling'' is a basic general problem in the studies of frustrated magnets including spin glasses \cite{chiral-SG-review}. The universality of the two transitions is also an interesting issue. However, the separation between $T_{\rm c}$ and $T_{\rm s}$ in the present system is extremely small, if any, which makes the solution of this problem technically difficult in spite of intensive studies done for more than two decades 
\cite{DHLee,TN,LKG,GN,Lee-Lee,Santiago-Jose,NGK,Olsson,Xu-Southern,Lee-Lee-98,Boubcheur-Diep,Ozeki-Ito,HPV05a,HPV05b,Thijssen,Grest89,Lee,Minnhagen85-FFXY,Granato-Dominguez,Korshunov,Olsso1n-Teitel,Capriotti,Lee-Lee-Kosterlitz,Lee-Granato-Kosterlitz,Choi-Stroud,Granato-Kosterlitz,Jeon-Park-Choi}.

 Under such circumstances, the purpose of the present paper is to clarify the nature of the phase transitions of the 2D FFXY model by performing extensive Monte Carlo (MC) simulations for systems as large as up to $L\sim O(10^{3})$. As was recently suggested by Hasenbusch, Pelissetto and Vicari (HPV) \cite{HPV05a,HPV05b}, such large system sizes are needed to extract asymptotic crucial properties of the system. Although our system size is comparable to Hsenbusch {\it et al\/}, we study various physical quantities including those which were not studied in Refs.\cite{HPV05a,HPV05b}. Namely, in addition to the Binder parameters, the correlation lengths and the helicity modulus studied in Refs.\cite{HPV05a,HPV05b}, we also investigate the specific heat, the chiral susceptibility and the vorticity modulus \cite{KK93,SX95} in order to shed further light on the ordering of the model. Furthermore, Refs.\cite{HPV05a,HPV05b} restricted their study just to the transition point, whereas we study an explicit temperature dependence of various physical quantities in a finite temperature range around the transition point(s).

 Then, we have found convincing evidence that the two phase transitions take place at mutually close but two distinct temperatures. We have also found that the critical properties of the chiral transition are well compatible with the Ising universality class. We corroborate Refs.\cite{HPV05a,HPV05b} in these conclusions. On the other hand, we have found that the universality of the spin transition is not compatible with the standard KT transition, a conclusion different from that of Refs.\cite{HPV05a,HPV05b}.

 The organization of the paper is as follows. In the next section, we introduce our model and explain the method of our simulations together with various physical observables we calculate. In sec. \ref{sec-results}, we present and discuss the results of our MC simulation. Finally in sec. \ref{sec-discussions}, we conclude with a summary and some additional discussions. In appendix \ref{sec-appendix}, some details of the coulomb-gas representation of the vorticity modulus are given.

\section{Model and Method}
\label{sec-model-methods}

 We study the fully frustrated {\it XY\/} (FFXY) model on a square lattice of size $N=L \times L$. We label the sites of the square lattice as $i=1,2,\ldots,N$ and denote their $x$ and $y$-coordinates as $x_{i}=1,2,\ldots,L$ and $y_{i}=1,2,\ldots,L$. Similarly, we label the plaquettes as $n=1,2,\ldots,N$ and denote their coordinates as $x_{n}=1,2,\ldots,L$ and $y_{n}=1,2,\ldots,L$.

The Hamiltonian of the model is given by,
\begin{equation}
H=-\sum_{\langle i,j \rangle}J_{ij} \vec{S}_{i}\cdot \vec{S}_{j} ,
\end{equation}
where $\vec S_i=(S_i^x, S_i^y)=(\cos \theta_i, \sin \theta_i)$ ($0\leq \theta_i <2 \pi$)  is the two-component spin variable at the site $i$. The summation $\langle ij \rangle$ is taken over all nearest-neighboring sites. The interaction bond takes two possible values $J_{ij}=\pm J$ such that every plaquette on the square lattice is frustrated , {\it i.e.\/}, the product of $J_{ij}$ around each plaquette is $\prod {\rm sign}(J_{ij})=-1$. In the following, we measure the (reduced) temperature $k_{\rm B}T/J$ in a unit with $k_{\rm B}/J=1$. We denote the thermal averages as $\langle \ldots \rangle$.

\subsection{Chirality-related observables}

 Let us introduce first physical observables associated with the chirality. We define the local chirality at a plaquette $n$ by,
\beq
\kappa_{n}=\frac{1}{2\sqrt{2}}\sum {\rm sign}(J_{ij})\sin (\theta_{i}-\theta_{j}),
\eeq
where the sum is taken over the four bonds around the plaquette. The normalization factor is chosen such that the absolute value of the chirality becomes unity in the ground state.
The checker-board-like order of the chirality can be detected though a  `staggered-chirality',
\beq
m_{\rm c} =\frac{1}{N}\sum_{n=1}^{N}(-1)^{x_{n}+y_{n}}\kappa_{n}.
\eeq

We study the onset of the chiral order via the chiral Binder parameter,
\beq
g_{\rm c}=\frac{3}{2}\left(1-\frac{\langle m_{\rm c}^{4}\rangle}{2\langle m_{\rm c}^{2}\rangle^{2}}\right) ,
\eeq
and the chiral correlation length,
\beq
\xi_{\rm c}=\frac{1}{2\sin(\pi/L)}\sqrt{\frac{ \langle| m_{\rm c}(\vec{0}) |^{2}\rangle}{\langle| m_{\rm c}(\vec{q}_{\rm min}) |^{2}\rangle}-1} ,
\eeq
where
\beq
m_{\rm c}(\vec{q})=\frac{1}{N}\sum_{n=1}^{N}(-1)^{x_{n}+y_{n}}e^{i \vec{q}\cdot \vec{r}_{n}}\kappa_{n}.
\eeq
is the Fourier transform of the staggered-chirality where $\vec{q}_{\rm min}= \frac{2\pi}{L}\vec{e}_{x}$ and $\vec r_n=(x_n,y_n)$.

In addition, we examine the staggered chiral susceptibility defined by,
\beq
\chi_{\rm c}=N\beta \langle m_{\rm c}^{2} \rangle, 
\eeq
where $\beta $ is the inverse temperature.

\subsection{Spin-related observables}

Next, we introduce the physical quantities associated with the spin. The spin order parameter is defined via the sublattice magnetization. The square lattice is decomposed into four sublattices, and the sublattice magnetization is defined by,
\beq
m_{{\rm s},\alpha}=\frac{1}{N/4} \left |\sum_{i \in {\rm sublattice}\ \alpha} \vec S_{i} \right|,
\eeq
where $\alpha=1,2,3,4$ denotes each sublattice.

In 2D,  the spin order parameter should vanish  in the thermodynamic limit $N\to \infty$ at any finite temperature. However, the spin may still establish a {\it quasi}-long-range order below a finite transition temperature $T_{\rm s}$. To detect such a spin order, we measure the spin Binder parameter,
\beq
g_{\rm s}=\frac{1}{4}\sum_{\alpha=1}^{4} \left( 2-\frac{\langle {m}_{{\rm s},\alpha}^{4}\rangle}{\langle {m}_{{\rm s},\alpha}^{2}\rangle^{2}} \right),
\eeq
and the spin correlation length,
\beq
\xi_{\rm s}=\frac{1}{4}\sum_{\alpha=1}^{4}\frac{1}{2\sin(\pi/L)}\sqrt{\frac{ \langle  m_{{\rm s},\alpha}(\vec{0}) ^{2}\rangle}{\langle  m_{{\rm s},\alpha}(\vec{q}_{\rm min})^{2}\rangle}-1} ,
\eeq
where
\beq
m_{{\rm s},\alpha}(\vec{q})=\frac{1}{N/4}\left | \sum_{i \in {\rm sublattice}\ \alpha} e^{i \vec{q}\cdot \vec{r_{i}}}
\vec{S}_{i} \right |,
\eeq
and $\vec{q}_{\rm min}=(2\pi /L, 0)$. We also calculate the sublattice spin susceptibility defined by,
\beq
\chi_{\rm s}=N\sum_{\alpha=1}^{4}\beta \langle m_{{\rm s},\alpha}^{2} \rangle.
\eeq

 In addition, we study the helicity modulus and the vorticity modulus which detect the rigidity of the system against appropriate spin deformations. The two quantities are related but have physically different meanings as explained below.

 The helicity modulus $\Upsilon$ measures the rigidity of the system against an infinitesimal uniform twist of the spins $\Delta \theta$ applied on the boundary. More precisely, it is defined through the change of the total free energy $\Delta F$ with respect to an infinitesimal twist on the spin configuration along, say,  $y$-axis $\theta_{i}-\theta_{i'} \to \theta_{i}-\theta_{i'}+(\Delta\theta /L$) where $i'$ is the nearest neighbour site of site $i$ along the $+y$ direction. One readily finds,
\beq
\Delta F=\Upsilon (\Delta \theta)^{2} + O(\Delta^{4}).
\eeq
The helicity modulus is then given by,
\begin{eqnarray}
 \Upsilon= \frac{1}{N} && \left [ \left \langle  \sum_{i} 
J_{ii'}\cos(\theta_{i}-\theta_{i'})\right \rangle \nonumber \right. \\
&& \left.  -\beta 
\left \langle \sum_{i} J_{ii'}\sin(\theta_{i}-\theta_{i'}) \right \rangle^{2}\right]. 
\end{eqnarray}

On the other hand, the vorticity modulus \cite{KK93,SX95} measures the  rigidity of the system against a creation of an extra vortex. Disregarding the discrete nature of the vorticity, we consider an {\it infinitesimal} increase of the vorticity represented by the phase differences $\theta_{i}-\theta_{j} \to \theta_{i}-\theta_{j}+m \phi_{ij}$ across each bond $<ij>$ where $m$ is assumed to be small. Here $\phi_{ij}=-\phi_{ji}$ is the solid angle between site $i$ and $j$ with respect to an arbitrarily chosen origin in the system where the core of the vortex is supposed to be located. In practice, we define the solid angle as $\phi_{ij}=\hat{e}_{ij} \cdot \hat{\phi}_{ij}/\sqrt{x_{i}^{2}+y_{i}^{2}}$. Here $\hat{e}_{ij}$ is the unit vector connecting the neighboring sites $i$ and $j$, and $\hat{\phi}_{ij}$ is a unit vector tangential to the circular path passing through the site $i$ enclosing the origin. One readily finds that the free-energy increase can be expressed as,
\beq
\Delta F= V m^{2} + O (m^{4}) ,
\eeq
with 
\begin{eqnarray}
 V= && 
\left \langle \sum_{<i,j>}  J_{ij} \phi^{2}_{ij} \cos(\theta_{i}-\theta_{j}) \right \rangle \nonumber  \\
&& 
-\beta \left \langle \sum_{<i,j>}  J_{ij} \phi_{ij} \sin(\theta_{i}-\theta_{j}) \right \rangle^{2}.
\end{eqnarray}
For large enough systems,  we expect the scaling behaviour,
\beq
V(T,L)\approx c(T)+v(T)\ln L,
\label{eq-vorticity-scaling}
\eeq
where $v(T)$ is the vorticity modulus.  In appendix \ref{sec-appendix}, we present a coulomb-gas representation of the vorticity modulus which might provide some insight into the physical meaning of this quantity.

In practice, we evaluate the vorticity modulus $v(T)$ in two different ways. The first way is to define the effective vorticity modulus by,
\beq
v_{1}(T,L)=\frac{V(T,L)}{\ln L} ,
\label{eq-vorticity-1}
\eeq
and examine its $L \to \infty$ limit. The second way is to define the effective vorticity modulus using the data at two different system sizes $L_{1}$ and $L_{2}$ \cite{SX95},
\beq
v_{2}(T,L_{1},L_{2})=\frac{V(T,L_{1})-V(T,L_{2})}{\ln (L_{1}/L_{2})} ,
\label{eq-vorticity-2}
\eeq
and examine its $L_{1},L_{2} \to \infty$ limit. The advantage of $v_{2}$ compared with the first one $v_{1}(T)$ is that the anticipated constant term $c(T)$ in
\eq{eq-vorticity-scaling} is cancelled so that a finite-size correction becomes smaller.

\subsection{Monte Calro simulations}

 We use a Monte Calro method which combines the Metropolis method and the over-relation method \cite{PP}. Within a single Monte Carlo Step (MCS), the spins are updated by one sweep using the Metropolis method which is followed by three consecutive over-relaxation sweeps. System sizes of $L=32,64,128,256,512,1024$ plus $L=384,768$ under periodic boundary conditions are studied. We use the histogram method to evaluate physical quantities around the critical temperature. Typical MC steps used for equilibration and measurements are $10^{8}$ MCS for $L=512,768,1024$ around the chiral transition temperature and $5\times 10^{6}$ MCS around the spin transition temperature.

\section{Results}
\label{sec-results}

 In this section, we present the results of our MC simulations.

\begin{figure}[t]
\includegraphics[width=0.45\textwidth]{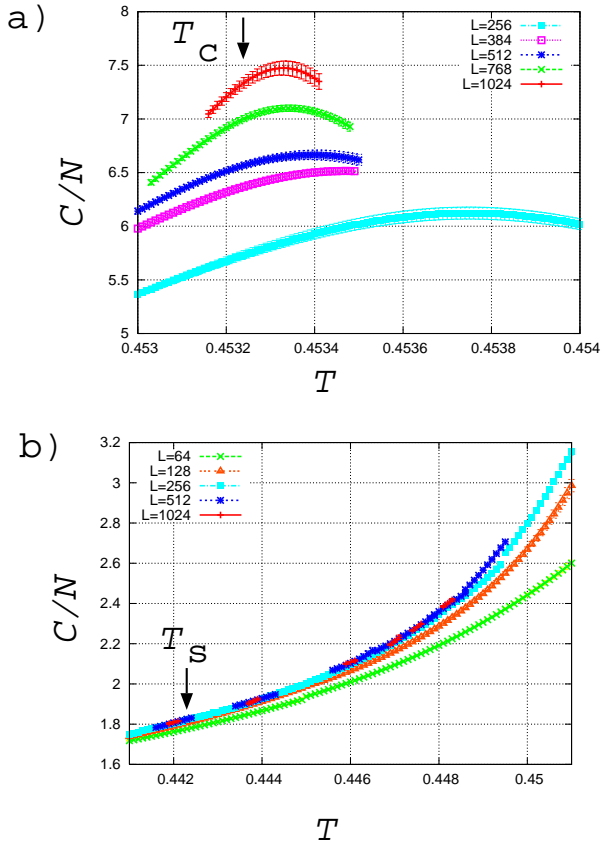}
\caption{The temperature dependence of the specific heat per spin. The panels a) and b) show the data around the chiral transition temperature $T_{\rm c}$ and spin transition temperature $T_{\rm s}$, respectively. The arrows indicate the estimated transition temperature for the chirality $T_{\rm c}$, and for the spin $T_s$.
}
\label{fig-heatcapacity}
\end{figure}

\begin{figure}[h]
\includegraphics[width=0.45\textwidth]{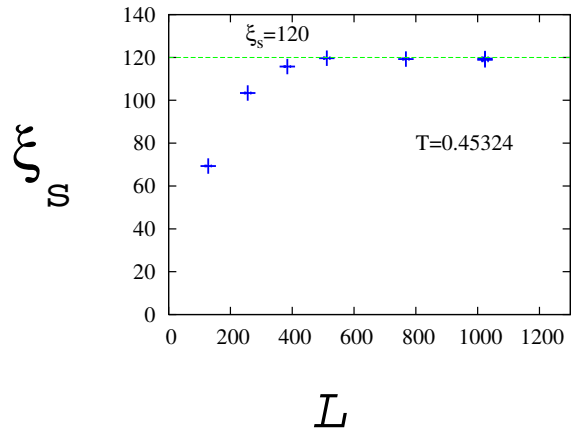}
\caption{
Spin correlation length $\xi_{s}(T,L)$ measured at the chiral transition temperature
$T_{\rm c}=0.45324$ for various system sizes.} 
\label{fig-raw-correlation-length-vs-L}
\end{figure}

\subsection{Specific heat}

\begin{figure}[h]
\includegraphics[width=0.45\textwidth]{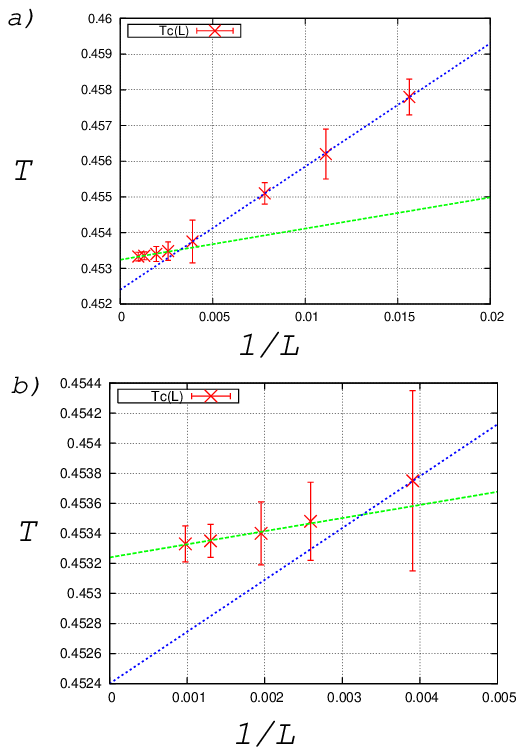}
\caption{The specific-heat-peak temperature $T_{\rm peak}(L)$ plotted versus the inverse linear size $1/L$, for (a) all sizes investigated, and for (b) larger sizes. The straight lines represent linear fits of the data at smaller sizes $L < 200$, and at larger sizes $L > 200$.
} 
\label{fig-heatcapacity-peak}
\end{figure}

 We begin with the specific heat $C=\beta^{2} (\langle H^{2} \rangle - \langle H \rangle^{2})$.  As shown in Fig.~\ref{fig-heatcapacity}(a), the specific heat exhibits a sharp peak as a function of the temperature at $T=T_{\rm peak}(L)$. The height of the peak increases with increasing $L$, suggesting the occurrence of a second-order phase transition.  Meanwhile, there is no indication  of other phase transitions. This could be explained if one recalls that the expected spin transition might be of the  KT-type which is known not to accompany a distinct anomaly in the specific heat. Therefore it is natural to expect that the specific-heat-peak is associated with the chiral transition.

 In order to estimate the bulk chiral transition temperature $T_c$, we show in Fig.~\ref{fig-heatcapacity-peak} the size dependence of the specific-heat-peak temperature $T_{\rm peak}(L)$. Interestingly, the system-size dependence of the peak temperature $T_{\rm peak}(L)$ exhibits a crossover behaviour around $L=L_\times \sim 200$: For sizes smaller than $L_\times \sim 200$, $T_{\rm peak}(L)$ exhibits a near-linear dependence, while for sizes greater than $L_\times$, $T_{\rm peak}(L)$ exhibits another near-linear dependence but with a different slope. By making a linear extrapolation of $T_{\rm peak}(L)$ for large enough system sizes $L > L_\times \sim 200$, we get an estimate of the chiral critical temperature, $T_{\rm c}=0.45324(1)$: See Fig.~\ref{fig-heatcapacity-peak}(b). It implies that the long-range order of the chirality is established at $T_{\rm c}=0.45324(1)$ via a second-order phase transition. Since the finite-size scaling ansatz implies  $T_{\rm peak}(L) \approx T_{\rm c}+{\rm const}L^{-1/\nu_c}$ where the critical exponent $\nu_c$ describes the divergence of the chiral correlation length, the chiral transition is characterized by $\nu_c\simeq 1$, which is consistent with the the Ising universality class. Our present estimate of $T_{\rm c}$ and the Ising nature of the transition are consistent with the results of many recent works, particularly with the most recent result by Hasenbusch et al who gave $T_{\rm c}=0.45324(1)$ \cite{HPV05a,HPV05b}.

 Now we turn to the issue of the crossover observed around $L_\times\sim 200$, which is likely to reflect a change in the nature of the ordering of the spin. In the inset of Fig.~\ref{fig-raw-correlation-length-vs-L}, we show the size dependence of the spin correlation length $\xi_s$ at the chiral transition point $T_{\rm c}=0.45324$ determined above. As can be seen from the figure, $\xi_{\rm s}(T_{\rm c})$ tends to saturate for large enough sizes of $L > 200$, yielding $\xi_{\rm s}(T_{\rm c})\sim 120$, which is indeed comparable to the crossover length scale $L_\times \sim 200$ determined above. This result indicates that the observed size-crossover reflects the change in the ordering behavior of the system from the `spin-chirality coupling' regime where the spin correlation dominates the ordering to the `spin-chirality decoupling' regime where the chiral correlation outgrows the spin correlation and dominates the ordering of the system.

 The universality of the chiral transition of the 2D FFXY model has been a matter of intense debate in the previous studies. Most probably, the source of the confusion is this size-crossover effect associated with the spin-chirality coupling/decoupling. If one looked at small enough systems compared with the crossover length $L_\times$, one should find an effective $\nu$-exponent close to unity as demonstrated in Fig.~\ref{fig-heatcapacity}(a), which would suggest the Ising universality. This was actually observed in simulations in early days \cite{MS84,Grest89}, where system sizes in the range $10 < L < 50$ were studied. However, if one focused on intermediate length scales around $L_\times$, one would find $\nu < 1$, which would suggest the non-Ising universality \cite{Santiago-Jose,Lee,Lee-Lee,Lee-Lee-98,Boubcheur-Diep,Ozeki-Ito}. Then, at length scales sufficiently larger than $L_\times$, the effective $\nu$-exponent would become closer to unity again \cite{HPV05a,HPV05b}. Our present estimates of $T_c$ and $\nu$ are in perfect agreement with the recent estimates of Hasenbusch {\it et al\/} \cite{HPV05a,HPV05b}.

To get more insight into the ordering process of the model, we proceed to analyze in the next subsections physical quantities which are directly related to the order parameters.

\begin{figure}[t]
\includegraphics[width=0.45\textwidth]{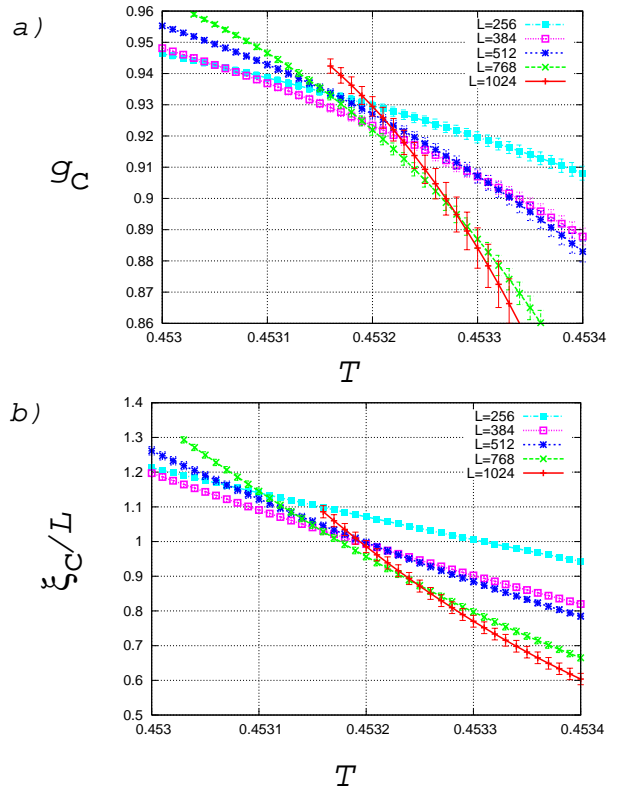}
\caption{
The temperature dependence of (a) the chiral Binder parameter $g_{\rm c}$, and of (b) the chiral correlation-length ratio $\xi_{\rm c}/L$.
}
\label{fig-binder-correlation-chirality}
\end{figure}
\begin{figure}[h]
\includegraphics[width=0.45\textwidth]{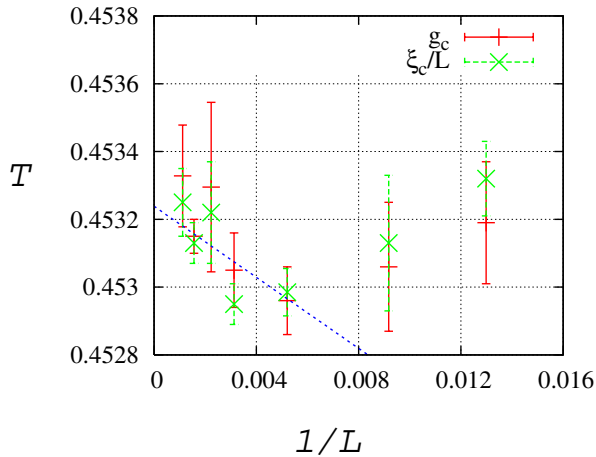}
\caption{Crossing temperatures of the chiral Binder parameter $g_{\rm c}$ and of the chiral correlation-length ratio $\xi_{\rm c}/L$ are plotted versus the inverse linear size $1/L$. The dashed line is a linear fit which yields $T_{\rm c}=0.45324(5)$.
} 
\label{fig-chirality-crossing}
\end{figure}

\subsection{Chirality-related quantities}

 In this subsection, we examine the ordering of the chirality. As shown in Fig.~\ref{fig-binder-correlation-chirality}(a), the chiral Binder parameter $g_{\rm c}(T,L)$ of different sizes exhibit clear crossings.  The scaled correlation-length, or the chiral correlation-length ratio $\xi_{\rm c}(T,L)$, also exhibits similar crossings as shown in Fig.~\ref{fig-binder-correlation-chirality}(b). 

To analyze the size dependence of the crossing points quantitatively, we label the system sizes in the ascending order $L_{1} < L_{2} \ldots$ and plot  in Fig.~\ref{fig-chirality-crossing} the crossings point $T_{\rm cross}(L)$ of the Binder parameter and of the chiral correlation-length ratio at adjacent sizes $L_{n+1}$ and $L_{n}$ as a function of the average size $L_{av}=\frac{L_{n+1}+L_{n}}{2}$.    
As in the case of the specific-heat-peak temperature, the crossing temperature $T_{\rm cross}(L)$ of both the Binder parameter and the correlation-length ratio exhibits a size-crossover  at around $L \sim 200$, the same crossover length observed in the specific heat, where a non-monotonic size dependence with an abrupt turnover takes place. With use of the data of larger sizes $L > L_\times$, the chiral transition temperature is estimated as $T_{\rm c}=0.45324(5)$, which agrees well with the estimate above obtained from the specific heat $T_{\rm c}=0.45324(1)$. 

\begin{figure}[t]
\includegraphics[width=0.45\textwidth]{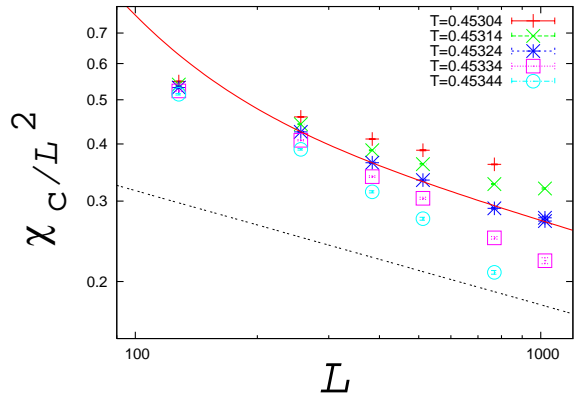}
\caption{
The chiral susceptibility divided by $L^2$ is plotted versus the system size $L$ on a log-log plot. 
The solid line represents a fit to the form $\chi_{\rm c}(T,L)=a(T)+b(T)L^{2-\eta_{\rm eff}(T)}$  of the data at $T=T_c=0.45324$. 
For comparison, a power-law $L^{-\eta}$ behavior with the Ising exponent $\eta=1/4$ is also shown by the dotted line.
}
\label{fig-chi}
\end{figure}
\begin{figure}[h]
\includegraphics[width=0.45\textwidth]{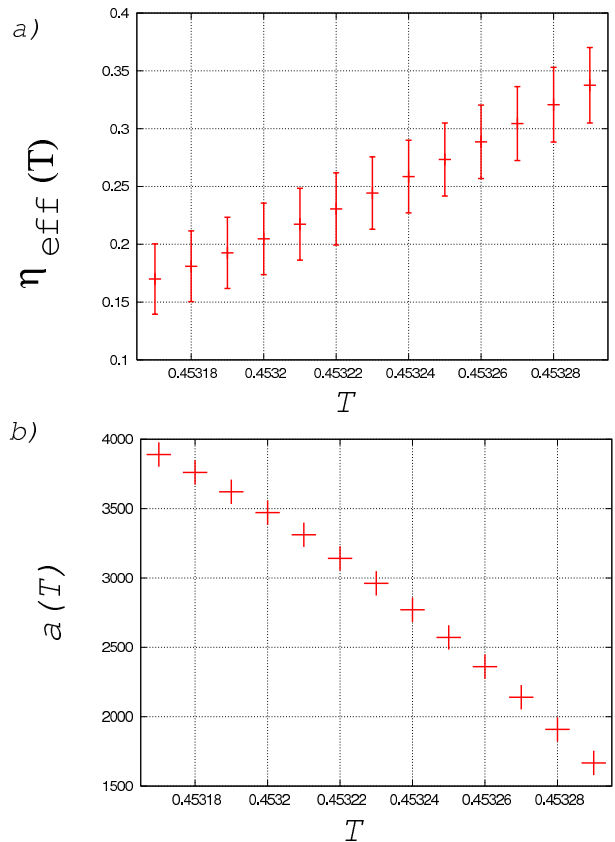}
\caption{
 The temperature dependence of the parameter values obtained in the fitting analysis of the chiral susceptibility based on the form $\chi_{\rm c}(T,L)=a(T)+b(T)L^{2-\eta_{\rm eff}(T)}$: a) the effective exponent $\eta_{\rm eff}(T)$, and b) the constant $a(T)$.
} 
\label{fig-eta}
\end{figure}

To examine the universality of the chiral transition further, we analyze the chiral susceptibility $\chi_{\rm c}$. At the chiral transition temperature $T_{\rm c}$, a power-law behaviour $\chi_{\rm c}(T_{\rm c},L) \propto L^{2-\eta_c}$  is expected with a chiral anomalous-dimension exponent $\eta_c$. We analyze the data  at each temperature $T$ shown in Fig.~\ref{fig-chi} by fitting them to the form $\chi_{\rm c}(T,L)=a(T)+  b(T)L^{2-\eta_{\rm eff}(T)}$ with three fitting parameters $a(T)$, $b(T)$ and $\eta_{\rm eff}(T)$. 

The temperature dependence of the effective exponent $\eta_{\rm eff}(T)$ is displayed in Fig.~\ref{fig-eta}(a). We find that, at the chiral transition temperature $T_{\rm c}=0.45324$ estimated above, the effective exponent $\eta_{\rm eff}(T)$ takes a value $0.25(3)$. This observation is consistent with the expected Ising value $\eta=1/4$, giving further support to the Ising nature of the chiral transition. 

In Fig.~\ref{fig-eta}(b), we display the temperature dependence of our another fitting parameter $a(T)$. Presumably, this parameter reflects fluctuations at short length scales where chiral and spin orders are not separated. The increase of $a(T)$ with decreasing temperature $T$ may be interpreted as due to the growth of the spin correlation length $\xi_{\rm s}(T)$ approaching $T_{\rm s}$.

\subsection{Spin-related quantities}

\begin{figure}[h]
\includegraphics[width=0.45\textwidth]{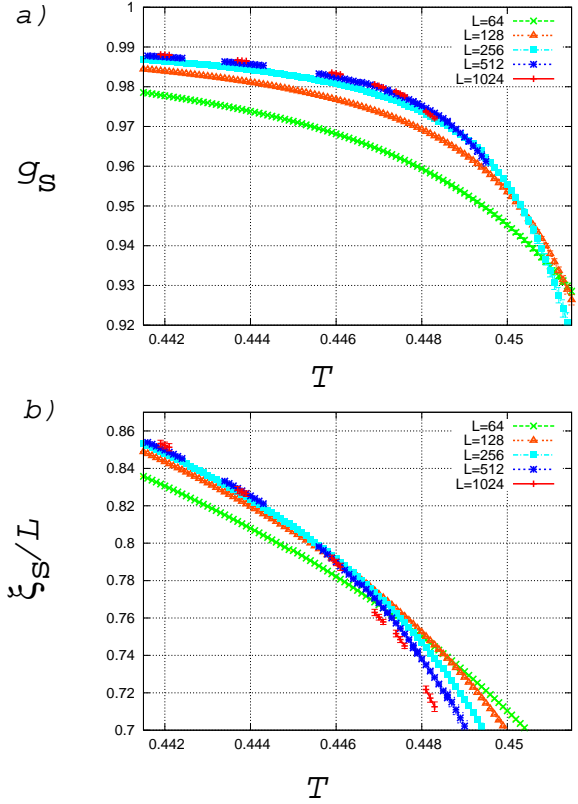}
\caption{
The temperature dependence of (a) the spin Binder parameter $g_{\rm s}$, and of (b) the spin correlation-length ratio $\xi_{\rm s}/L$. Evidently the crossing temperatures are lower than the chiral transition temperature $T_{\rm c}=0.45324$.
} 
\label{fig-binder-correlation-spin}
\end{figure}

\begin{figure}[h]
\includegraphics[width=0.45\textwidth]{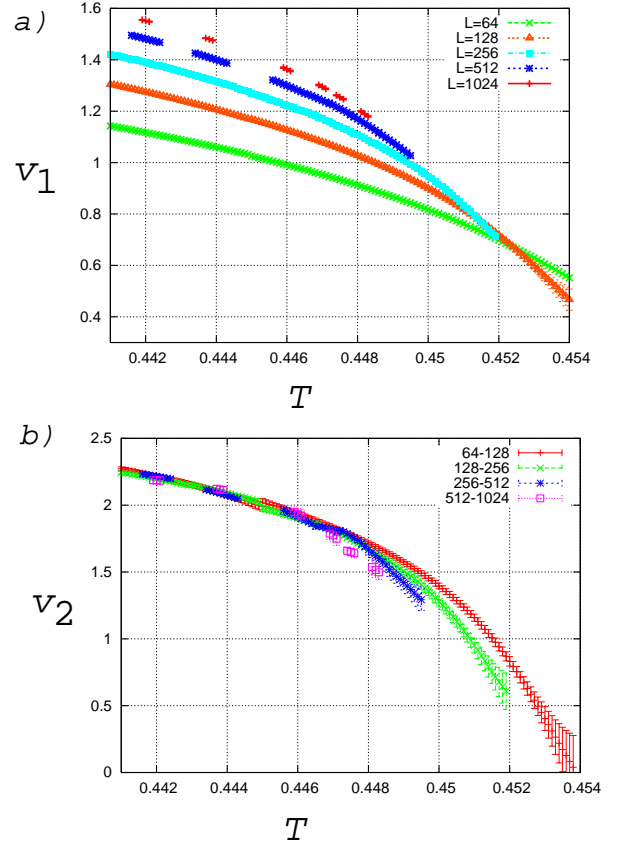}
\caption{
The temperature dependence of the two types of the effective vorticity modulus, (a) $v_{1}$ defined by \eq{eq-vorticity-1}, and b) $v_{2}$ defined by \eq{eq-vorticity-2}.
} 
\label{fig-vorticity}
\end{figure}

Next, we examine the ordering of the spin. The spin Binder parameter $g_{\rm s}(T,L)$ and the spin correlation-length ratio $\xi_{\rm s}(T,L)/L$ are shown in Fig.~\ref{fig-binder-correlation-spin}(a) and (b), respectively. Data of different sizes exhibit crossings as in the case of chirality shown in Fig.~\ref{fig-binder-correlation-chirality}. From the raw data, it is already evident that the spin crossing temperatures are considerably lower than chiral transition temperature $T_{\rm c}=0.45324(1)$. It means that the chirality and the spin exhibit two separate transitions.

In addition, the raw data also suggests a qualitative difference in the nature of the orderings of the spin and the chirality.  In case of the spin shown in Fig.~\ref{fig-binder-correlation-spin}, the separation between the data points of different sizes tends to become increasingly small for larger systems, exhibiting a marginal behavior, whereas  in case of the chirality shown in Fig.~\ref{fig-binder-correlation-chirality} the data points of different system sizes remain well separated even for larger systems. This difference reflects the fact that, while the chirality exhibits a finite long-range order at finite temperature even in 2D, the spin cannot establish a true long-range order at finite temperature but only a quasi-long-range order.

The effective vorticity modulus $v_{1}$ and $v_{2}$ defined by \eq{eq-vorticity-1} and \eq{eq-vorticity-2} are displayed in Fig.~\ref{fig-vorticity}(a) and (b), respectively. The effective vorticity modulus $v_{1}$ exhibits a  stronger finite-size effect compared with $v_{2}$, as expected. $v_{2}$ exhibit merging behaviour at low temperatures suggesting the onset of the quasi-long ranged order. Again, it can be seen that the crossing temperatures of the vorticity modulus are considerably lower than the chiral transition temperature $T_{\rm c}=0.45324(1)$.

\begin{figure}[h]
\includegraphics[width=0.45\textwidth]{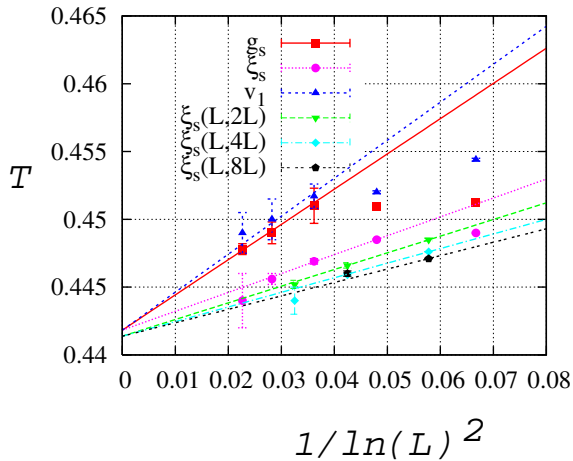}
\caption{Crossing temperatures of the spin Binder parameter $g_{\rm s}$  (Fig.~\ref{fig-binder-correlation-spin}(a)), of the spin correlation-length ratio $\xi_{\rm s}/L$  (Fig.~\ref{fig-binder-correlation-spin}(b)), and of the effective vorticity modulus $v_{1}$ (Fig.~\ref{fig-vorticity})(a)) are plotted versus  $(1/\ln L)^{2}$.  $L$ in the horizontal axis is the average system size. The combined straight-line fits for three kinds of crossing temperatures assuming a common $T_{\rm s}$ yields $T_{\rm s}=0.4418(5)$.
In addition, crossing temperatures of the spin correlation-length ratio $\xi_{\rm s}/L$
between the two sizes $L$ and $sL$ with $s=2,4,8$ are also shown as a function of $1/(\ln L)^{2}$. The combined fit using the last three data sets assuming a common $T_{\rm s}$ yields $T_{\rm s}=0.4414(3)$.
} 
\label{fig-spin-crossing}
\end{figure}

 Now we estimate quantitatively the spin transition temperature $T_s$ based on our data of the crossing temperatures of several observables presented above. Thus, the crossing temperatures $T_{\rm cross}(L)$ of the spin Binder parameter $g_{\rm s}(T,L)$, the spin correlation-length ratio $\xi_{\rm s}(T,L)/L$ and the effective vorticity modulus $v_{1}$ between the two adjacent sizes $L_1$ and $L_2$ are plotted against $1/(\ln L_{av})^{2}$ in Fig.\ref{fig-spin-crossing}.  Note that this form is motivated by the standard KT form $\xi_s \sim \exp[c/\sqrt{T-T_s}]$. From the combined fit of these data  where a common spin transition temperature $T_{\rm s}$ is assumed, we get an estimate of the bulk spin transition temperature $T_{\rm s}=0.4418(5)$.

In Fig.\ref{fig-spin-crossing} we also display the size dependence of the crossing points of the spin correlation-length ratio $\xi_{\rm s}(T,L)/L$ 
between the two sizes $L$ and $sL$ with $s=2,4,8$ 
plotted as a function of $1/(\ln L)^{2}$ . From the combined fit of these three data sets where a common spin transition temperature $T_{\rm s}$ is assumed, we get $T_{\rm s}=0.4414(3)$ which is consistent with our estimate presented above.

\begin{figure}[t]
\includegraphics[width=0.45\textwidth]{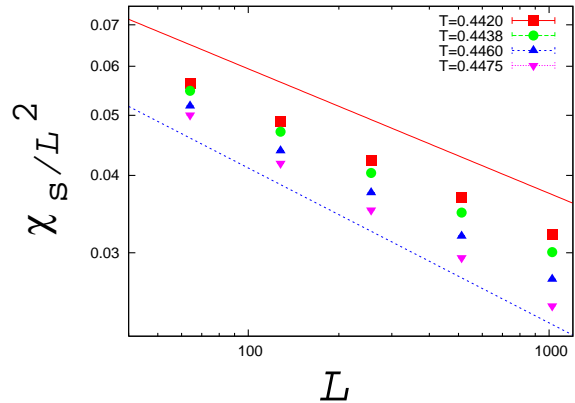}
\caption{
The sublattice spin susceptibility divided by $L^2$ is plotted versus $L$ on a log-log plot. The straight lines represent power laws $L^{-\eta}$ with $\eta=1/4$ (green broken line) and with $\eta=0.201$ (red solid line).
}
\label{fig-chi-spin}
\end{figure}

In order to examine the universality of the spin transition, we also analyze the sublattice spin susceptibility $\chi_{\rm s}$. At and below the spin transition temperature $T_{\rm s}$, a power-law behaviour $\chi_{\rm s}(T_{\rm c},L) \propto L^{2-\eta(T)}$  is expected with a spin anomalous-dimension exponent $\eta(T)$. We analyze the data shown in Fig.~\ref{fig-chi-spin} by fitting them to the form $\chi_{\rm s}(T,L)=b(T)L^{2-\eta_{\rm eff}(T)}$ with two fitting parameters $b(T)$ and $\eta_{\rm eff}(T)$ at each temperature $T$. 

\begin{figure}[h]
\includegraphics[width=0.45\textwidth]{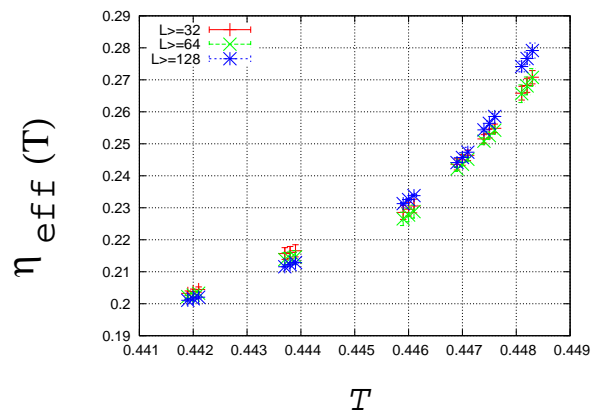}
\caption{
The temperature dependence of the effective spin exponent $\eta_{\rm eff}(T)$ extracted from the fitting of the sublattice spin susceptibility $\chi_{\rm s}(T,L)$ to the form $\chi_{\rm s}(T,L)=b(T)L^{2-\eta_{\rm eff}(T)}$ using the data of $L \geq 32$, $\geq 64$ and $\geq 128$. 
} 
\label{fig-eta-spin}
\end{figure}

The temperature dependence of the effective exponent $\eta_{\rm eff}(T)$ is displayed in Fig.~\ref{fig-eta-spin}(a). We find that, at the spin transition temperature $T_{\rm s}=0.4418(5)$ estimated above, the effective exponent $\eta_{\rm eff}(T)$ takes a value $0.201(2)$.  Remarkably, this value is significantly smaller than the value of the conventional Kosterlitz-Thouless (KT) transition  $\eta=1/4$ \cite{KT}. This observation suggests that the universality of the spin-transition of the present 2D FFXY model might be different from that of the the conventional KT transition.

\begin{figure}[t]
\includegraphics[width=0.45\textwidth]{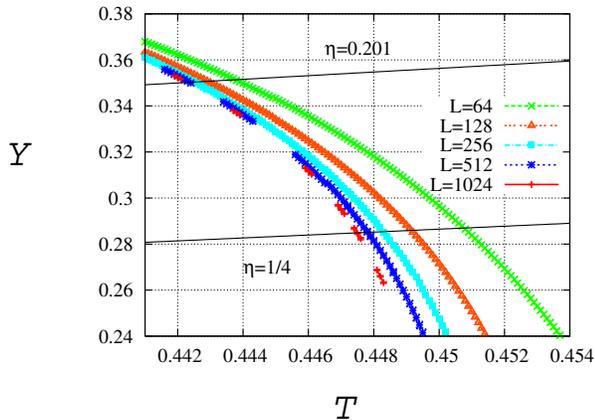}
\caption{
  The temperature dependence of the helicity modulus. The straight lines represent $T/(2\pi\eta)$ with $\eta=0.201$ and $1/4$.
} 
\label{fig-helicity}
\end{figure}

\begin{figure}[h]
\includegraphics[width=0.45\textwidth]{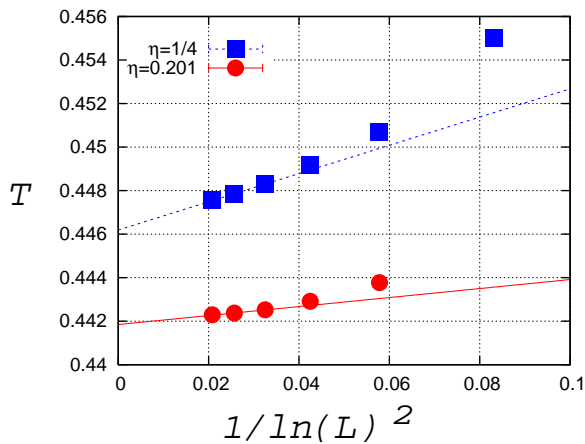}
\caption{
Intersection points of the helicity modulus obtained by assuming $\eta(T_{\rm s})=0.201$ and $1/4$ are plotted versus $1/(\ln L)^{2}$. The straight lines represent the linear fits of the data.
} 
\label{fig-spin-crossing-helicity}
\end{figure}

To get further insight, we also examine the helicity modulus. In 2D {\it XY\/} models, the helicity modulus is expected to exhibit a discontinuous jump from zero at higher temperatures to a nonzero value $T_{\rm s}/(2\pi \eta(T_{\rm s}))$ at the critical temperature $T_{\rm s}$. From the raw data displayed in Fig.~\ref{fig-helicity}, we extract the intersection points between the helicity-modulus data of each size and a straight line $T/(2\pi \eta(T_{\rm s}))$ by setting $\eta(T_{\rm s})=0.201(2)$ as obtained above. Then, the size dependence of the resulting intersection points is displayed in Fig.~\ref{fig-spin-crossing-helicity}. Assuming that the correction term scales as $1/\ln^{2}(L)$, we find $T_{\rm s}=0.44185(5)$. This value of $T_s$ is consistent with our estimate above $T_{\rm s}=0.4418(5)$ obtained from the Binder parameter, the correlation-length ratio and the vorticity modulus. 

 If one assumes instead the conventional KT value $\eta(T_{\rm s})=1/4$ in the above analysis of the helicity modulus as was done in Refs.\cite{Olsson,HPV05a,HPV05b}, one obtains $T_{\rm s}=0.44619(4)$ as displayed in Fig.~\ref{fig-spin-crossing-helicity}. The latter value agrees well with the estimate of $T_{\rm s}=0.446$ (Ref.\cite{Olsson}) and $T_{\rm s}=0.4461(1)$ (Ref.\cite{HPV05a,HPV05b}).  However, this value is not compatible with our present estimate of $T_{\rm s}$ obtained from other quantities {\it without assuming a specific value for the exponent $\eta(T_{\rm s})$\/}, $T_{\rm s}=0.4415(3)$. 

 Our result indicates that the jump of the helicity-modulus is greater than the so-called `universal jump' for the conventional KT transition \cite{NK77,Jose-Kadanoff-Kirkpatrick-Nelson,Minnhagen-review}. Lee and Lee \cite{Lee-Lee-98} made a similar observation in the FFXY model on the triangular lattice. Indeed, such a possibility was mentioned occasionally since early days of research \cite{TJ83}.



\section{Summary and discussions}
\label{sec-discussions}

 We studied the ordering of the spin and the chirality of the 2D FFXY model on the square lattice by extensive Monte Carlo simulations performed up to very large system size to circumvent crossover effects. We found the the chiral transition takes place at $T_{\rm c}=0.45324(1)$, while spin transition takes place at $T_{\rm s}=0.4418(5)$. Thus, the model certainly exhibits the spin-chirality decoupling, where $T_{\rm s}$ lies below  $T_{\rm c}$ by about 2.6\%.

\begin{figure}[h]
\includegraphics[width=0.45\textwidth]{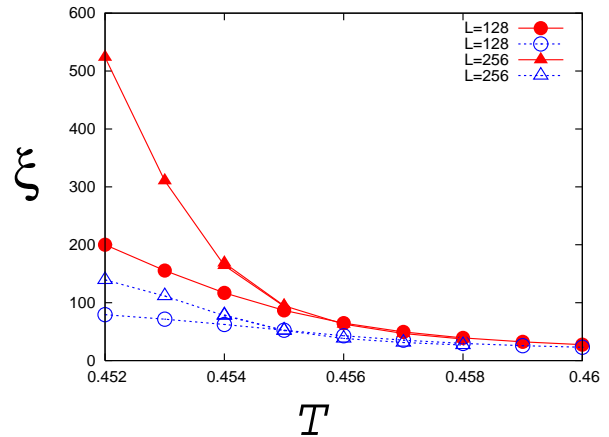}
\caption{
Direct comparison of the chiral and the spin correlation lengths for the sizes $L=128$ and 256. The full/open symbols are the finite-size data of the chiral/spin correlation length, respectively. 
} 
\label{fig-raw-correlation-length}
\end{figure}

 As pointed out by Hasenbusch {\it et al\/} \cite{HPV05a,HPV05b},  in order to observe separation of the two transitions it is crucial to simulate large enough system sizes. In Fig.~\ref{fig-raw-correlation-length} we display the raw data of the correlation length of the chirality and of the spin. Evidently, the two correlation lengths are almost equal at higher temperatures $T > 0.468$, behaving side by side. This is the spin-chirality coupling behavior expected in the high-temperature regime. At lower temperatures, the bulk chiral correlation length becomes definitely longer than the bulk spin correlation length. For instance, at $T=0.456$ we find $\xi_{\rm c} \sim 65$ while $\xi_{\rm s} \sim 45$. In the temperature range  $T < 0.458$, we get into the spin-chirality decoupling regime. At the chiral transition temperature $T_{\rm c}=0.45324(1)$, the chiral correlation length diverges while the spin correlation length $\xi_{\rm s}(T_{\rm c},L)$ stays at a finite value around $120$  as shown in the inset of Fig.~\ref{fig-raw-correlation-length}.

 The universality class of each transition was also studied. The universality class of the chiral transition is well compatible with the Ising universality. On the other hand, the universality class of the spin transition is different from the conventional KT transition. 

Possibility of the non-universal jump of the helicity modulus was pointed out on the general ground of a refined renormalization-group scheme by Minnhagen \cite{Minnhagen85} and was argued to be relevant to the fully frustrated XY model \cite{Minnhagen85-FFXY}. However the true connection to the frustration system was not necessarily clear. The spin and the chirality orderings of the fully frustrated XY model was also studied by Choi and Stroud by means of a renormalization group analysis of the coupled XY model which is believed to be essentially equivalent to the FFXY model \cite{Choi-Stroud}. The possible non-universal jump of the helicity modulus was suggested \cite{Granato-Kosterlitz,Jeon-Park-Choi}. The latter RG scheme \cite{Choi-Stroud,Granato-Kosterlitz,Jeon-Park-Choi} is quite appealing since it aimed at studying the ordering behaviours of both the chirality and the spin in a unified manner. However, the assumption underlying the derivation of the RG equations, i.~e. the smallness of the fugacity of the {\it both} the usual spin vorticies and extra charges responsible for the ordering of the chirality, each being justifiable in the low-temperature and in the high-temperature limits, cannot be satisfied asymptotically upon renormalization at any temperature. It will be very interesting to investigate further the nature of the spin transition in the presence of the chiral long-ranged order, which is not considered in conventional theories for the KT transition.

The numerical calculations were carried out partly on SX8 at Yukawa Institute of Theoretical Physics in Kyoto University and on SX-8R at the Cybermedia Center in Osaka University. This work is supported by Grant--Aid for Scientific Research on Priority Areas ``Novel States of Matter Induced by Frustration'' (19052006).

\appendix
\section{Vorticity modulus in the coulomb gas representation}
\label{sec-appendix}

In order to get some insight into the meaning of the vorticity modulus, we derive in this appendix its coulomb representation.

The partition function of our model can be represented as,
\beq
Z= \int \prod_{i} d\theta_i \prod_{<ij>}e^{-\beta E_{ij}(\{\theta_{ij}\},\{A_{ij}\})} ,
\eeq
with
\beq
E_{ij}(\{\theta_{ij}\},\{A_{ij}\})= -J \cos(\theta_{i}-\theta_j-A_{ij}) ,
\eeq
The parameter $A_{ij}$ (vector potential) is understood as an anti-symmetric tensor,
\beq
A_{ij}=-A_{ji}.
\eeq
For a ferromagnetic/antiferromagnetic bond, we choose $A_{ij}=0/\pi$, respectively.

\begin{figure}[h]
\begin{center}
\includegraphics[width=0.45\columnwidth]{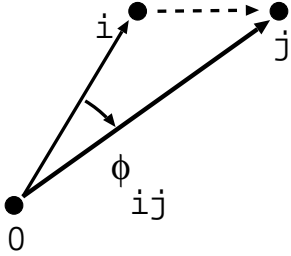}
\end{center}
\caption{Solid angle $\phi_{ij}$ between sites $i$ and $j$ with respect to the origin O.}
\label{fig-phi}
\end{figure}

We consider the change in the free energy by an {\it infinitesimal} increase of the vorticity $m$,
\beq
V=\left. \frac{\partial^{2} F(m)}{\partial m^{2}}\right|_{m=0}, 
\eeq
where $F(m)$ is the free-energy of the system with a modified vector potential,
\beq
A_{ij} \to A_{ij} + m \phi_{ij},
\label{eq-additional-vectorpotential}
\eeq
where $\phi_{ij}$ is an `angle' between the sites $i$ and $j$ with respect to the origin O, as shown in Fig.~\ref{fig-phi}. We set,
\beq
\phi_{ij}= - \phi_{ji}.
\eeq

By the standard steps of mappings \cite{Jose-Kadanoff-Kirkpatrick-Nelson}, 
we find that the partition function of the system can be represented as,
\beq
Z= Z_{\rm sw} \prod_{\vec{n}}\sum_{q_{\vec{n}=-\infty}}^{\infty}
e^{-\beta {\cal H}(\{\tilde{q}_{\vec{n}}\})},
\eeq
where $Z_{\rm SW}$ is the contribution from spin-waves and ${\cal H}(\{\tilde{q}_{\vec{n}}\})$ is an effective charge Hamiltonian for the charge variables $\tilde{q}_{\vec{n}}$ defined on the dual lattice with its lattice points $\vec{n}=(x_{n},y_{n})$ located at the centers of the plaquettes of the original lattice,
\beq
\tilde{q}_{\vec{n}}=q_{\vec{n}}+b_{\vec{n}}.
\eeq
Here $q_{\vec{n}}$ is an integer and $b_{\vec{n}}$  is a `flux' defined by
\beq
b_{\vec{n}}=\frac{1}{2\pi}\sum_{<ij> \hspace*{.1cm}\rm{around} \hspace*{.1cm}\vec{n}} A_{ij},
\eeq
where the sum is taken around the plaquette at  $\vec{n}$ in a counter-clockwise direction.

One can see that the contribution from the additional term $m \phi_{ij}$ in the vector potential \eq{eq-additional-vectorpotential} to the flux $b_{\vec{n}}$ is zero almost everywhere except at the origin $\vec{n}=\vec{O}$;
\beq
b_{\vec{n}}=m \delta_{\vec{n},\vec{0}}+\frac{1}{2\pi}\sum_{<ij> \hspace*{.1cm}\rm{around} \hspace*{.1cm}\vec{n}} A_{ij} .
\eeq
Here it is made evident that $m$ can be interpreted indeed as an {\it infinitesimal} charge or vorticity.

 We then find,
\begin{eqnarray}
&& \left. \frac{\partial^{2} (-\beta F)}{\partial m^{2}}\right |_{m=0}
 = \left \langle 
\frac{\partial^{2} (-\beta {\cal H})}{\partial (\tilde{q}_{\vec{0}})^{2}} 
\right \rangle_{\rm c} \nonumber \\
&+&  \left \langle \left ( \frac{ \partial (-\beta {\cal H})}{\partial (\tilde{q}_{\vec{0}})} 
 \right)^{2}  
\right \rangle_{\rm c}
-  \left \langle 
\frac{\partial (-\beta {\cal H})}{\partial (\tilde{q}_{\vec{0}})} 
\right \rangle^{2}_{\rm c} ,
\end{eqnarray}
where 
\beq
\langle \ldots \rangle_{\rm c}  \equiv 
\frac{\prod_{\vec{n}}\sum_{q_{\vec{n}=-\infty}}^{\infty} \ldots
e^{-\beta {\cal H}(\{\tilde{q}\})}}{\prod_{\vec{n}}\sum_{q_{\vec{n}=-\infty}}^{\infty}
e^{-\beta {\cal H}(\{\tilde{q}\})}} .
\eeq
An explicit form of the charge Hamiltonian ${\cal H}(\{\tilde{q}\})$ can be written as,
\beqa
{\cal H}(\{\tilde{q}\})
= \sum_{\vec{n}_{1},\vec{n}_{2}} g_{2}(\vec{n}_{1},\vec{n}_{2}) 
\tilde{q}_{\vec{n}_{1}}\tilde{q}_{\vec{n}_{2}}+\ldots
\eeqa
where higher-body ({\it i.e.\/}, $4$-,$6$-,\ldots body) terms are neglected for simplicity, which  corresponds to the Villain's approximation. The two-body interaction $g_2(\vec{n}_{1},\vec{n}_{2})$ is essentially the same as the spin-wave Green's function, 
\begin{eqnarray}
g_2(\vec{n}_{1},\vec{n}_{2})&\equiv &\sum_{\vec{k}}\frac{e^{i\vec{k}\cdot(\vec{n}_{1}-\vec{n}_{2})}}{4-2(\cos(k_{x})+\cos(k_{y}))} \nonumber \\
& \simeq & \frac{1}{2\pi}\int_{\pi/L}^{\pi}dk k \frac{e^{ik|n_{1}-n_{2}|}}{k^{2}}.
\end{eqnarray}
In the first equation, the summation over all possible wavevectors $\vec{k}=(k_{x},k_{y})$ allowed in the periodic system of size $L \times L$ is taken.  The Green's function behaves at large distances as,
\begin{equation}
g_2(\vec{n}_{1},\vec{n}_{2}) \approx 2\pi\ln (|\vec{n}_{1}-\vec{n}_{2}|) + 2\pi \ln (2 \sqrt{2} e^{\gamma}),
\end{equation}
where $\gamma$ is Euler's constant while 
\begin{equation}
g_2(\vec{n},\vec{n}) \approx 2\pi \ln (L) + 2\pi \ln (2 \sqrt{2} e^{\gamma}),
\end{equation}
which is the bare vortex core energy of a finite sized system of size $L$.

 Within the Villain's approximation, we find,
\begin{eqnarray}
&&  V=\left. \frac{\partial^{2} F}{\partial m^{2}}\right |_{m=0}  \nonumber \\ 
&&  = g_{2}(\vec{0},\vec{0}) 
  -\beta \sum_{\vec{n},\vec{n'}}
g_2(\vec{0},\vec{n})g_2(\vec{0},\vec{n'})
(\langle \tilde{q}_{\vec{n}}\tilde{q}_{\vec{n'}}
\rangle _{\rm c}
-\langle \tilde{q}_{\vec{n}} \rangle_{\rm c}
\langle \tilde{q}_{\vec{n'}} \rangle_{\rm c}
).\nonumber \\
\label{eq-V-charge}
\end{eqnarray}
The first term on the r.h.s of the last equation is obviously the vortex creation energy which generally scales as $c(0)+v(0)\ln(L)$ with a `bare' vorticity modulus at zero temperature $v(0)=2\pi$ and a constant $c(0)=2\pi \ln (2 \sqrt{2} e^{\gamma})$.  The second term would be interpreted as a `screening' term due to thermal fluctuations of other charges. 
 On general grounds, we speculate that the same scaling holds at finite temperatures below the critical temperature $T_{\rm s}$,
\beq
V=c(T)+v(T)\ln(L), \qquad T < T_{\rm s}, 
\eeq
with a temperature-dependent vorticity modulus $v(T)$. Above $T_{\rm s}$, we  have $v(T)=0$ in the thermodynamic limit $L \to \infty$.

The obtained expression of $V$, \eq{eq-V-charge}, should be compared with the corresponding expression for the helicity modulus which is given by \cite{Minnhagen-review,Olsson},
\begin{equation}
\Upsilon = 1-\frac{\beta}{N}
\sum_{\vec{n},\vec{n'}}
g_{2}(\vec{n},\vec{n'})
(\langle \tilde{q}_{\vec{n}}\tilde{q}_{\vec{n'}}
\rangle _{\rm c}
-\langle \tilde{q}_{\vec{n}} \rangle_{\rm c}
\langle \tilde{q}_{\vec{n'}} \rangle_{\rm c}).
\end{equation}

\end{document}